\documentclass{article}

\usepackage{graphicx}
\usepackage[absolute]{textpos} 
\usepackage{calc} 
\textblockorigin{25.4mm}{25.5mm}%

\usepackage{pdfpages}

\usepackage{xcolor}
\definecolor{OrangeGSSI}{RGB}{237,113,45}
\definecolor{GrayDISIM}{RGB}{138,140,138}
\definecolor{BlueDISIM}{RGB}{31,70,93}
\definecolor{BordeauxUNIVAQ}{RGB}{123,13,25}
\definecolor{DarkBlue}{RGB}{25,25,112}

\usepackage[pdfpagemode={UseOutlines},bookmarks=true,bookmarksopen=true,
   bookmarksopenlevel=0,bookmarksnumbered=true,hypertexnames=false,
   colorlinks,linkcolor={DarkBlue},citecolor={DarkBlue},urlcolor={DarkBlue},
   pdfstartview={FitV},unicode,breaklinks=true]{hyperref}

\usepackage{enumerate}
\usepackage{setspace}
\usepackage[square, numbers, comma, sort&compress]{natbib}

\usepackage{amssymb}

\usepackage{footnote}
\makesavenoteenv{table}
\makesavenoteenv{tabular}
   
\DeclareFontFamily{OT1}{ams}{}
\DeclareFontShape{OT1}{ams}{m}{n}{ <-> msam10 }{}
\DeclareFontShape{OT1}{ams}{m}{it}{ <-> msam10 }{}
\DeclareFontShape{OT1}{ams}{bx}{n}{ <-> msbm10 }{}
\DeclareFontShape{OT1}{ams}{bx}{it}{ <-> msbm10 }{}

\newcommand*{\typename}[1]{\def\Type{#1}}
\newcommand*{\titlename}[1]{\def\Title{#1}}
\newcommand*{\versionnumber}[1]{\def\Version{#1}}
\newcommand*{\versiondate}[1]{\def\Date{#1}}

\newcommand*{\firstauthorname}[1]{\def\FirstAuthor{#1}}
\newcommand*{\secondauthorname}[1]{\def\SecondAuthor{#1}}
\newcommand*{\thirdauthorname}[1]{\def\ThirdAuthor{#1}}
\newcommand*{\fourthauthorname}[1]{\def\FourthAuthor{#1}}

\usepackage{ifthen}
\usepackage{footnote}
\usepackage{amssymb}
\newboolean{showcomments}
\setboolean{showcomments}{true} 
\ifthenelse{\boolean{showcomments}}
{\newcommand{\nb}[2]{
		\fcolorbox{gray}{yellow}{\bfseries\sffamily\scriptsize#1}
		{\sf\small$\blacktriangleright$\textit{#2}$\blacktriangleleft$}
	}
	
}
{\newcommand{\nb}[2]{}
	
}

\usepackage{vmargin}
\usepackage{balance}
\usepackage[normalem]{ulem} 
\usepackage{listings}
\usepackage{AMMALanguages}
\usepackage{times}
\usepackage{dsfont}
\usepackage{algorithm}
\usepackage{wrapfig}
\usepackage{mdwmath}
\usepackage{mdwtab}
\usepackage{amsmath}
\usepackage{url}
\usepackage{xcolor}
\usepackage{paralist}
\setpapersize{A4}
\setmarginsrb  			{ 1.0in}  
                        	{ 0.6in}  
                        	{ 1.0in}  
                        	{ 0.8in}  
                        	{  20pt}  
                        	{0.25in}  
                        	{   9pt}  
                        	{ 0.3in}  

\newcommand{\HRule}{\rule{159.2mm}{0.5mm}} 

\typename{Protocol for a Systematic Mapping Study on}
\titlename{Collaborative Model-Driven Software Engineering}
\versionnumber{0.1}
\versiondate{January 27, 2015}
\firstauthorname{\href{mailto:mirco.franzago@graduate.univaq.it}
			    {\textcolor{BlueDISIM}{Mirco Franzago $^{\diamond}$}}}
\secondauthorname{\href{mailto:ivano.malavolta@gssi.infn.it}
			    {\textcolor{black}{Ivano Malavolta $^{\star}$}}	}
\thirdauthorname{\href{mailto:henry.muccini@univaq.it}
				 {\textcolor{BlueDISIM}{Henry Muccini $^{\diamond}$}}}
\fourthauthorname{\href{mailto:davide.diruscio@univaq.it}
				 {\textcolor{black}{Davide Di Ruscio $^{\diamond}$}}}

\title{\Type~\Title}

\begin{document}

\begin{titlepage}

\includepdf[pages={1},scale=.8]{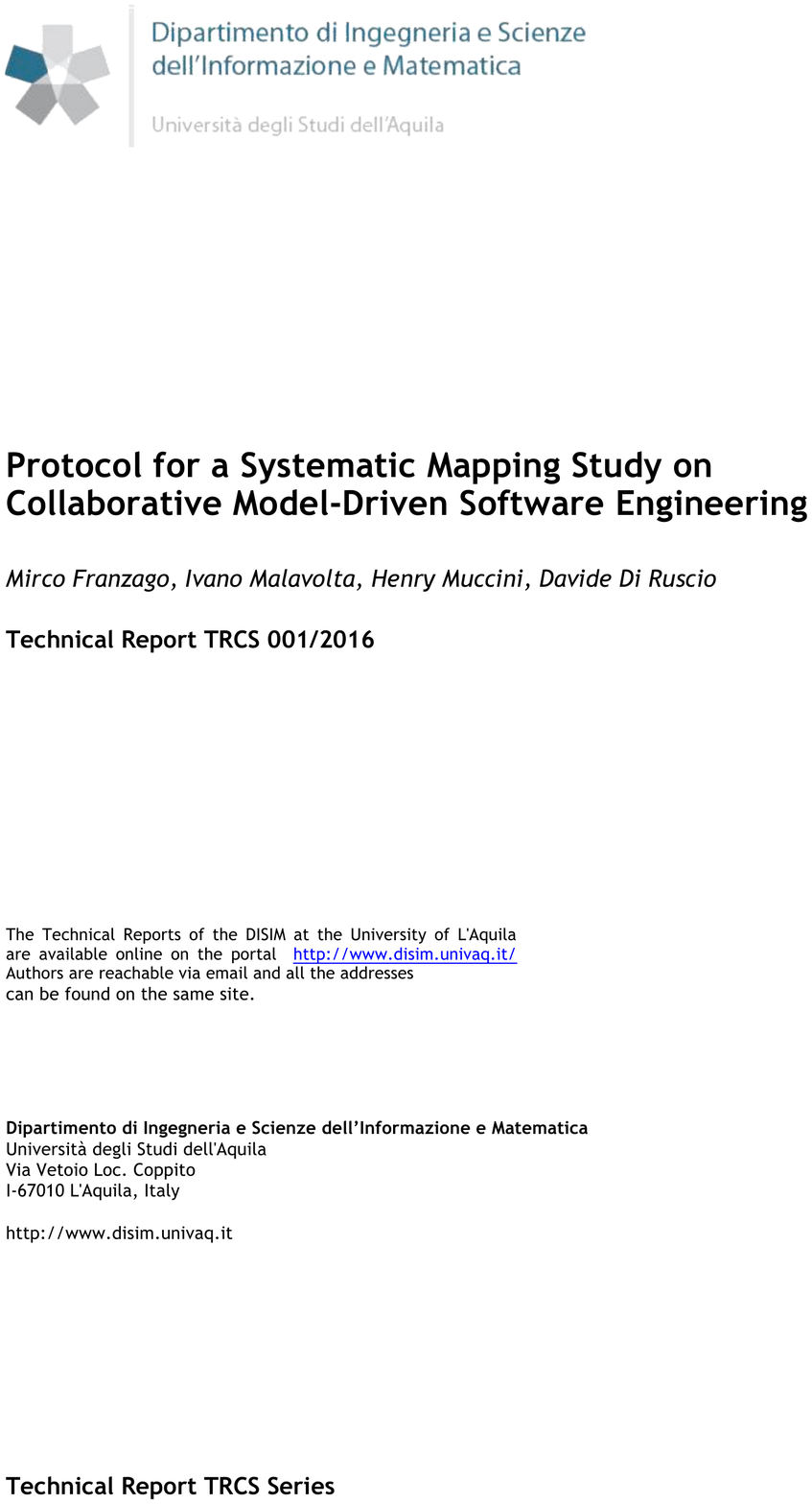}

\begin{center}
\begin{textblock*}{60mm}(1.0in,0.6in+20pt)
	  \href{http://cs.gssi.infn.it}{\includegraphics[width=60mm]{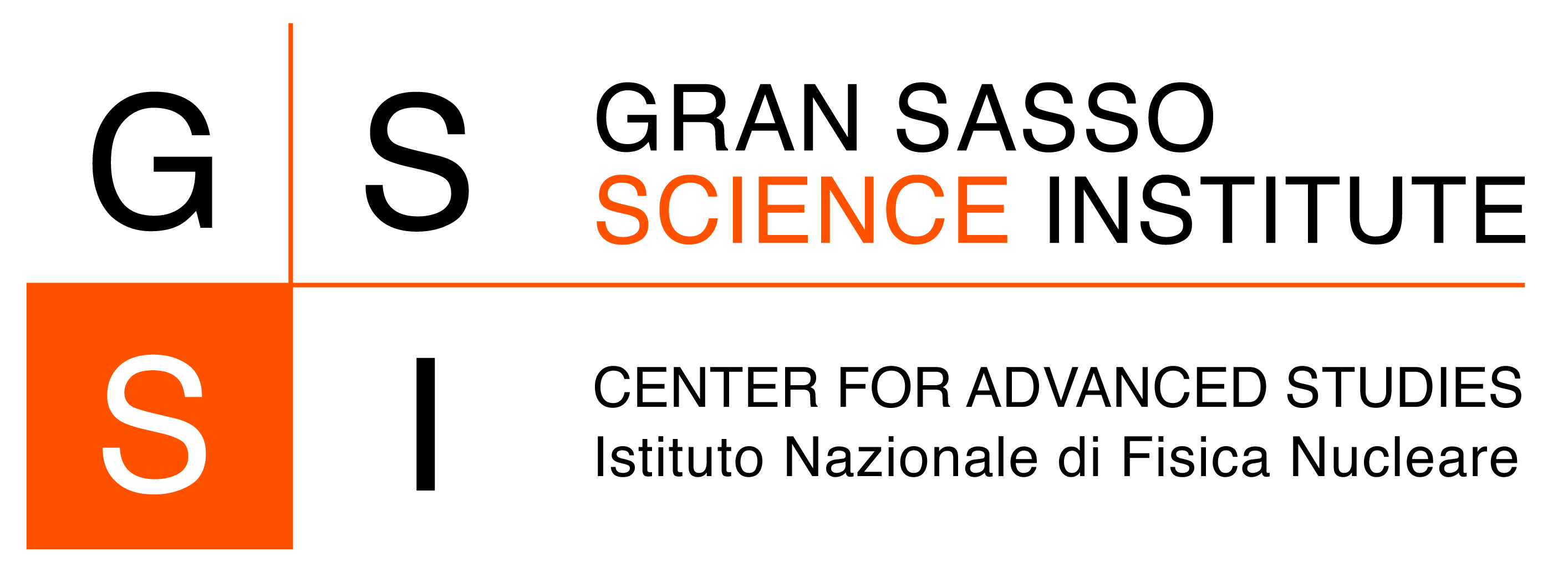}}
\end{textblock*}%
\begin{textblock*}{84.80mm}(1.0in+74.4mm,0.6in+20pt)%
	  \href{http://www.disim.univaq.it/main/index.php}{\includegraphics[width=84.80mm]{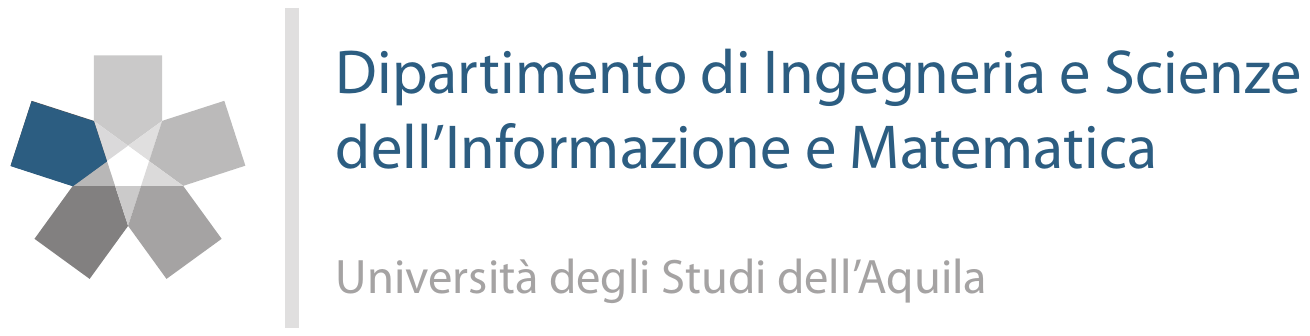}} 
\end{textblock*}%
\begin{textblock*}{159.2mm}(1.0in,2.8in)
       {\Large \sc \Type} \\[1mm] 
       \noindent\textcolor{GrayDISIM}{\HRule} \\[5mm]
       {\huge \bfseries \Title} \\[3mm] 
       \noindent\textcolor{GrayDISIM}{\HRule} \\[5mm]
\end{textblock*}%
\begin{textblock*}{146.5mm}(1.5in,6in)
       \begin{flushleft}
       {\Large {\sc	
       			\FirstAuthor		\\[2mm] 
       			\SecondAuthor	\\[2mm]
				\ThirdAuthor		\\[2mm] 
				\FourthAuthor	\\[2mm] }}
       \end{flushleft}\end{textblock*}%
\begin{textblock*}{159.2mm}(1.0in,9in)
       {{\Large $^{\diamond}$ \href{http://www.univaq.it/en/index.php}{\bf \textcolor{BordeauxUNIVAQ}{University of L'Aquila}} \\[1.5mm]
         \Large \href{https://goo.gl/maps/IS7Lx}{\textcolor{black}{Via Giovanni Di Vincenzo 16/B - 67100 L'Aquila - Italy}} }} \\[1mm]
\end{textblock*}%
\begin{textblock*}{159.2mm}(1.0in,10in)
       {\Large $^{\star}$ \href{http://www.gssi.infn.it/index.php/en/}{\bf \textcolor{black}{GS}\textcolor{OrangeGSSI}{S}\textcolor{black}{I}
       \textcolor{black}{Gran Sasso} \textcolor{OrangeGSSI}{Science} \textcolor{black}{Institute}} \\[1.5mm]
        \Large \href{https://goo.gl/maps/9Cj77}{\textcolor{black}{Viale Francesco Crispi, 7 - 67100 L'Aquila - Italy}} } \\[1mm]
\end{textblock*}%
\begin{textblock*}{159.2mm}(1.0in,10.6in)
   \href{http://cs.gssi.infn.it/people/zacchialun/}{\textcolor{black}{Document style $\copyright$ Yuriy Zachia Lun}} \\[1mm]
\end{textblock*}%
\end{center}
\end{titlepage}
\null\newpage
\setstretch{1.1}
\cleardoublepage
\pagenumbering{roman}
\begin{center}
{\sc \Type} \\[-2mm]
\noindent\textcolor{GrayDISIM}{\rule{150mm}{0.2mm}} \\[1mm]
{\Large \bf \Title} \\[-2mm]
\noindent\textcolor{GrayDISIM}{\rule{150mm}{0.2mm}} \\[1mm]
\end{center}
\vspace{5mm}
{\sc Abstract}\\[1mm]
This document describes the review protocol describing the conduct of a systematic mapping study on collaborative model-driven software engineering.\\[5mm]
{\sc Keywords}\\[1mm]
Systematic Mapping Study, Collaborative Modeling, Model-driven Engineering \\[5mm]
\pagebreak
\tableofcontents
\listoffigures
\listoftables
\pagebreak
\cleardoublepage
\pagenumbering{arabic}
\section{Background and rationale}\label{sec:background}
%


\noindent \textbf{Context.} Collaborative software engineering (CoSE) deals with methods, processes and tools for enhancing collaboration, communication, and co-ordination (3C) among team members \cite{collaborativeSEBookChapter19}. The importance of CoSE is
evidenced in a number of sources \cite{collaborativeSEBook,Kolovos:2013:RRT:2487766.2487768,brambillaBook_MDEinPractice} and recently empowered by the prominence of agile methods, open source software projects, and global software development \cite{collaborativeSEBookChapter19}.
CoSE is not only about software development team members, but it can embrace also external and non-technical stakeholders, like customers and final users, as
advised by current research on Participatory Design methods \cite{Schuler:1993:PDP:563076,Vredenburg:2002:SUD:503376.503460}.

When focusing on software design, considered to be one of the key aspects of software engineering~\cite{bruegge2004object}, multiple stakeholders with different technical knowledge and background collaborate on the development of the system~\cite{designThinking}.
In this context, shared (abstract) \textit{models} of system are extremely valuable since they allow each stakeholder to focus on
domain-specific concepts and to abstract upon the aspects of the system in which she is more expert. 
A model is {\em ``a reduced representation of some system that highlights the properties of interest from a given viewpoint''}\cite{Selic:2003:PMD:942589.942714},
and it is a specific design artifact that can be either graphical, XML-based, or textual. Models have an unequivocally defined semantics, which allow precise information exchange and many additional usages.
Modeling, as opposed to simply drawing, grants a huge set of additional advantages and uses, including: syntactical validation, model analysis, model simulation, model transformations, model execution (either through code generation or model interpretation), and model debugging \cite{brambillaBook_MDEinPractice}.

Nowadays, collaborative modeling performed by multiple stakeholders is gaining a growing interest in both academia and practice~\cite{Kolovos:2013:RRT:2487766.2487768,brambillaBook_MDEinPractice}.
However, it poses a set of research challenges, such as large and complex models management,
support for multi-user modeling environments, and synchronization mechanisms like models migration and merging, conflicts management, models versioning and rollback support~\cite{Kolovos:2013:RRT:2487766.2487768}. A body of knowledge in the scientific literature about collaborative model-driven software engineering (MDSE) exists. Still, those studies are scattered across different independent research areas, such as software engineering, model-driven engineering languages and systems, model integrated computing, etc., and a study classifying and comparing the various approaches and methods for collaborative MDSE is still missing.

Under this perspective, a systematic mapping study (SMS)~\cite{wohlin2012experimentation} can help researchers and practitioners in (i) having a complete, comprehensive and valid picture of the state of the art about collaborative MDSE, and (ii) identifying potential gaps in current research and future research directions.


\noindent \textbf{Goal of this work.} We are interested in identifying and classifying approaches, methods and techniques that support collaborative MDSE. We focus on those approaches in which several distributed technical and/or non-technical stakeholders collaborate to produce models of a software system, working in a shared environment, either synchronously or asynchronously. Stakeholders can include, but are not limited to, technical figures (modelers, designers, developers), domain experts, non-technical managers, customers, and users of the software system. We are interested in identifying and analyzing the different approaches to support multi-user modeling tasks where the design models can be either domain-specific or domain-independent. In any case, studied approaches must consider the models as first class elements within the whole design process. Also, studied approaches must provide synchronization mechanisms, e.g. conflicts management/resolution, conflicts avoidance, versioning and rollback support.


\subsection{Existing systematic studies on the topic}\label{sec:existingslr}


Based on our knowledge and after a manual search, we did not find any existing systematic study on the topic. In any case, in the following we report those studies that, even if they have different scopes and objectives, it is interesting to compare our research in order to better understand:
\begin{inparaenum}[(i)]
	\item our research focus and
	\item how other studies are not sufficient to answer the research questions of our study.
\end{inparaenum}

Table \ref{tab:existingslr} shows the existing systematic studies, their focus and quality assessment. Based on the criteria explained in \cite{Kitchenham2010792}, we calculate the total score of each study by summing up the answer to each specific question Q1-Q4 (Yes(Y)=1, Partly(P)=0.5, No(N)=0):

\begin{enumerate}[{Q}1)]
	\setlength{\itemsep}{-3pt}
	\item Are the systematic study's inclusion and exclusion criteria described and appropriate?
	\item Is the literature search likely to have covered all relevant studies?
	\item Did the reviewers assess the quality/validity of the included studies?
	\item Were the basic data/studies adequately described?
\end{enumerate}

\begin{table}[h]
	\begin{center}
		\begin{tabular}{| p{1cm} | p{1cm} | p{0.7cm} | p{0.7cm} | p{0.7cm} | p{0.7cm} |p{1cm} | p{4.5cm} |}
			\hline
			\textbf{Study} & \textbf{Year} & \textbf{Q1} & \textbf{Q2} & \textbf{Q3} & \textbf{Q4} & \textbf{Total score} & \textbf{Focus} \\ \hline
			\cite{collaborationModels_SR} & 2011 & Y & Y & Y & Y & 4.0 & Ways of collaboration used in DSD \\
			~\cite{toolsUsedInGSE_SMR} & 2012 & Y & P & P & Y & 3.0 & Tools supporting distributed teams in GSE \\
			~\cite{challengesInCollaborativeModeling_LR} & 2008 & P & Y & P & P & 2.5 & Collaborative modeling support systems  \\
			\hline
		\end{tabular}
		\caption{Existing systematic studies on collaborative software engineering or collaborative modeling}
		\label{tab:existingslr}
	\end{center}
\end{table}


A systematic literature review on the models of collaboration in the domain of distributed software development (DSD) is presented in \cite{collaborationModels_SR}. This study focuses on the models and tools for DSD \textit{based on life cycle of traditional software development (and its variations), where each phase of the cycle is performed}. Differently, in our study we focus on the various aspects of the collaboration in the model-driven software engineering domain, where the models are placed as first-class artifacts.


In \cite{toolsUsedInGSE_SMR} a systematic mapping study is proposed with the objective to discover all the tools available in the literature supporting global software engineering (GSE) activities.
Our work will be more comprehensive because it will identify (in addition to tool support) also methods, techniques and approaches to support modeling activities in collaborative settings with the focus on design activities. Moreover, in the literature there are other systematic studies in the DSD and GSE scope \cite{DSD_tertiary}, but all of them focus on tools and/or approaches to address issues like collaboration process management, team members awareness or collaboration tools support;
there is no existing study specifically focusing on collaborative MDSE.

Finally, in \cite{challengesInCollaborativeModeling_LR} the authors identify challenges and best practices in collaborative modeling activity, where modelers, end-users and experts are all involved in the model-based design of the system, and collaborate to create a shared understanding of the system under development (or part of it). Among others, the main difference between this study and ours is that in \cite{challengesInCollaborativeModeling_LR} collaborative modeling is considered as \textit{the joint creation of a shared graphical representation of a system}, i.e., a sketching activity where the created models are used as communication means between team members. This kind of models (possibly conforming to some syntactical rules) are very different from our definition of models, where modeling is a complex activity based on precise models whose semantic is rigorously defined according to a specific modeling language. These models allow precise information exchange but also many additional usages, including: syntactical validation, model checking, model transformation, code generation \cite[$\S$~2.1]{brambillaBook_MDEinPractice}.

\subsection{The need for a systematic mapping study on the topic}\label{sec:need}

The need for a systematic mapping study on collaborative MDSE has been introduced in Section~\ref{sec:background}.
This research complements the existing studies described in Section~\ref{sec:existingslr} to investigate the state-of-research about collaborative MDSE (cf. Table~\ref{tab:existingslr}).

So far, a large body of knowledge has been proposed in both modeling software systems (e.g., model-driven engineering techniques, domain-specific modeling languages, model transformations, etc.), and collaboration for software production (e.g., global software engineering methods, methods for participatory design of software systems, version control systems, etc.).
Even if the progress of research on the above mentioned areas has started more than a decade ago and the various research communities are still very active, we did not find any evidence that could help us in assessing the impact of existing research on \textit{collaborative MDSE}.
Thus, in this systematic work we aim to identify, classify, and understand existing research on collaborative MDSE. Those activities will help researchers and practitioners in identifying trends, limitations, and gaps of current research on collaborative MDSE and its future potential.
\section{Research implementation}

\subsection{Process}

\begin{figure}
	\centering
	\includegraphics[width=\columnwidth]{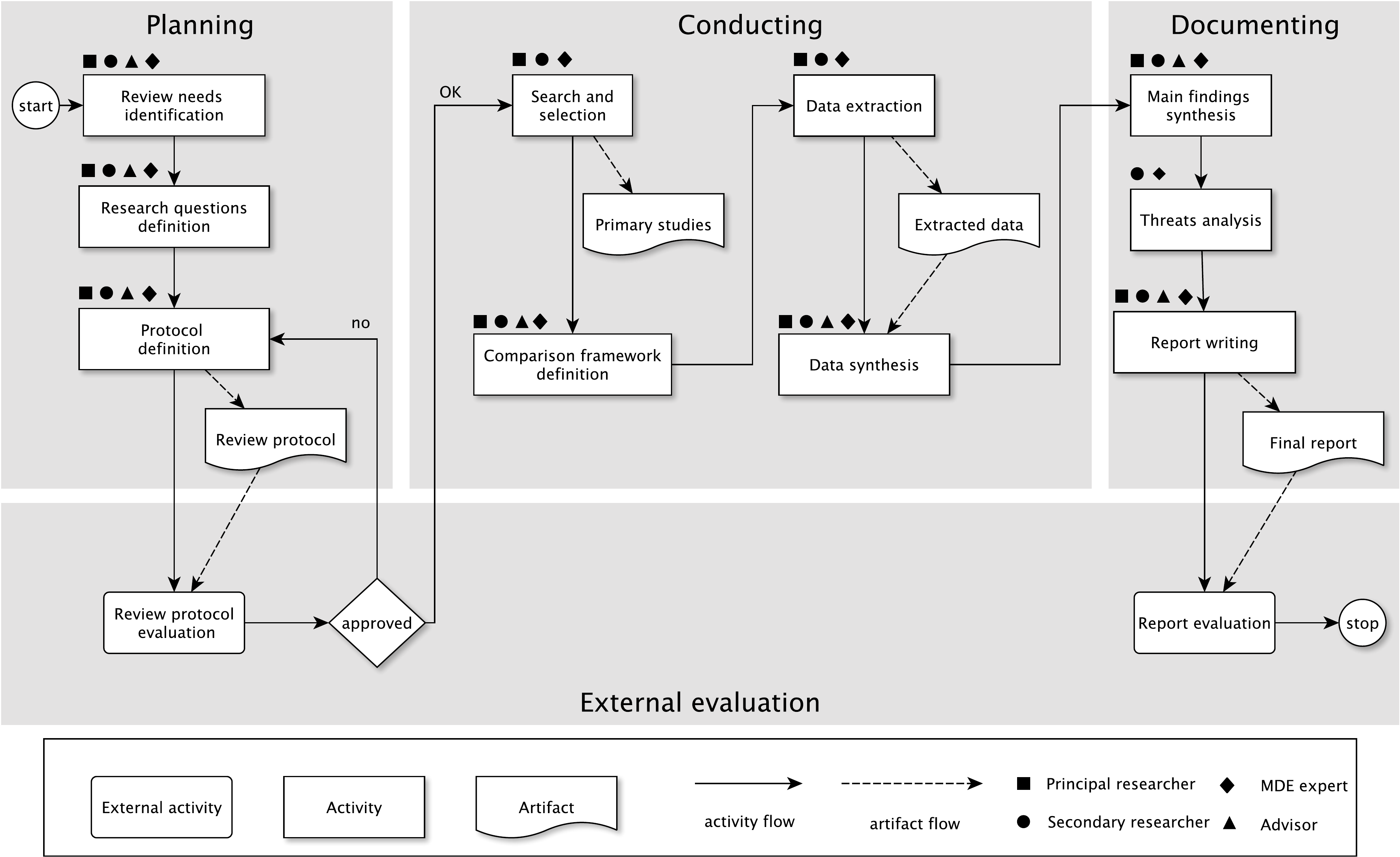}
	\caption{Overview of the whole review process}
	\label{fig:process}
\end{figure}

Figure~\ref{fig:process} shows the overview of the whole process of our research.
The overall process can be divided into three main phases, which are the classical ones for carrying on a systematic study~\cite{kitchenham2007guidelines,wohlin2012experimentation}: planning, conducting, and documenting.
Each phase can have a number of output artifacts, e.g., the planning phase produces the review protocol described in this document. In order to mitigate potential threats to validity and possible biases, review protocol and final reports artifacts will be circulated to external experts for independent review. More specifically, we identified two classes of external experts: SLR experts and domain experts. SLR experts are contacted for getting feedback about the proposed review protocol, possible unidentified threats to validity, possible problems in the overall construction of the review; whereas, domain experts are contacted for getting feedback about whether the proposed review protocol and final reports, can be effective with respect to the object of our mapping study (i.e., collaborative MDSE). Table~\ref{tab:reviewers} shows the external reviewers we contacted and replied. External reviewers had a time span of ten days to provide their feedback about the proposed artifacts.

\begin{center}
	\begin{table}[htbp]\center
		\begin{tabular}{ p{4cm} | p{4cm} }
			\hline
			\textbf{SLR experts} & \textbf{Domain experts} \\ \hline
			Muhammad Ali Babar & Dimitrios Kolovos \\
			Patricia Lago & \\
			\hline
		\end{tabular}
		\caption{Contacted external reviewers}
		\label{tab:reviewers}
	\end{table}
\end{center}

In the following we will go through each phase of the process, highlighting its main activities and produced artifacts.

\subsubsection{Planning}\label{sec:planning}
Planning is the first step of our review process, and it aims at (i) establishing the need for performing a mapping study on collaborative MDSE (see Section~\ref{sec:need}), (ii) identifying the main research questions (see Section~\ref{sec:rq}), and (iii) defining the protocol to be followed by the involved researchers while carrying on each step of the whole review process (see the remainder of this document). The output of our planning phase is a well-defined review protocol, that actually is this document itself. The produced review protocol will undergo an external evaluation by the previously identified SLR- and domain-experts.

\subsubsection{Conducting}\label{sec:conducting}
In this phase we will set the previously defined protocol into practice. More specifically,
we will perform the following activities:

\begin{itemize}
	\item \textit{Search and selection}: we will (i) consider the search strings identified in Section~\ref{sec:search} and we will apply them to electronic data sources, and (ii) apply the backward and forward snowballing techniques for expanding the set of considered studies. The output of this activity is a comprehensive list of all the candidate entries; each entry will have a set of metadata associated with it (e.g, title, authors, etc.).
	Duplicated entries will identified and merged by matching them by \texttt{title}, \texttt{authors}, \texttt{year}, and venue of \texttt{publication}. 
	Then, the potentially relevant studies identified in the previous activity will be filtered in order to obtain the final list of primary studies to be considered in later activities of the protocol. Section~\ref{sec:search} will describe in details the search strategy of this research.
	\item \textit{Comparison framework definition}: in this activity we will define the set of parameters that will be used to compare the primary studies. The main outcome of this activity is the data extraction form, which is designed to collect the information needed for analyzing the primary studies. The data extraction form will be designed based on the research questions~\cite{wohlin2012experimentation}.
	\item \textit{Data extraction}: In this activity we will go into the details of each primary study, and we will fill a corresponding data extraction form, as defined in the previous activity. Filled forms will be collected and aggregated in order to be ready to be analyzed during the next activity. More details about this activity will be presented in Section~\ref{sec:extraction}.
	\item \textit{Data synthesis}: this activity will focus on a comprehensive summary and analysis of the data extracted in the previous activity. The main goal of this activity is to elaborate on the extracted data in order to address each research question of our study (see Section~\ref{sec:rq}). This activity will involve both quantitative and qualitative analysis of the extracted data. The details about this activity are in Section~\ref{sec:analysis}.
\end{itemize}

\subsubsection{Documenting}\label{sec:documenting}
This phase is fundamental for reasoning on the obtained findings and for evaluating the quality of the systematic  study. The main activities performed in this phase are: (i) a thorough elaboration on the data extracted in the previous phase with the main aim at setting the obtained results in their context from both the academic and practitioners point of view, (ii) the analysis of possible threats to validity, specially to the ones identified during the definition of the review protocol (in this activity also new threats to validity may emerge), and (iii) the writing of a set of reports describing the performed mapping study to different audiences (see Section~\ref{sec:audience}). Firstly, produced reports will be evaluated by SLR- and domain- experts; secondly,
some of them will be submitted to scientific journals, conferences and magazines for professionals, thus they will undergo also a peer reviewed evaluation by the community.

\subsection{Team}
Four researchers will carry on this study, because a 'too small' team size (e.g., single reviewer) may have difficulties in controlling potential biases~\cite{tertiaryAli}. Each researcher has a specific role within the team; these are identified roles:
\begin{enumerate}[-]
	\item \textit{Principle researcher}: PhD student with knowledge about model-driven engineering and development; he will perform the majority of activities from planning the study to reporting;
	\item \textit{Secondary researcher}: post-doctoral researcher with expertise in both SLR methodologies and model-driven engineering; he is mainly involved in (i) the planning phase of the study, and (ii) supporting the principle researcher during the whole study, e.g., by reviewing the data extraction form, selected primary studies, extracted data, produced reports, etc.;
	\item \textit{MDE expert}: senior researcher with a several-years expertise on model-driven engineering methods and techniques; he is mainly involved in supporting the principle and secondary researchers with respect to any potential issue or discussion related to the MDE methodology;
	\item \textit{Advisor}: senior researcher with many-years expertise in software engineering and model-based design and development. He makes final decision on conflicts and options to 'avoid endless discussions'~\cite{tertiaryAli}, and supports other researchers during the data synthesis, findings synthesis, and report writing activities.
\end{enumerate}

\section{Research questions}\label{sec:rq}
Specifying the research questions is a most crucial part of doing systematic mapping study~\cite{Brereton2007571}.
Before going into the details of the identified research questions, we formulate the goal of this research by using the Goal-Question-Metric perspectives (i.e., purpose, issue, object, viewpoint~\cite{gqm}). Table \ref{tab:gqm}
shows the result of the above mentioned formulation.

\begin{center}
	\begin{table}[h]
		\begin{tabular}{ p{2cm} | p{11cm} }
			\textit{Purpose} & Identify, classify, and understand \\
			\textit{Issue} & the publication trends, characteristics, and challenges \\
			\textit{Object} & of existing collaborative MDSE approaches  \\
			\textit{Viewpoint} & from a researcher's viewpoint. \\
		\end{tabular}
		\caption{Goal of this research}
		\label{tab:gqm}
	\end{table}
\end{center}

In the following we present the research questions we translated from the above mentioned overall goal. For each research question we also provide its primary objective of investigation. The research questions are:

\begin{itemize}[-]

  \item \textbf{RQ1}: \textit{What are the characteristics of collaborative MDSE approaches?}
This research question has been decomposed into more detailed sub-questions in order for it to be addressed. Those sub-questions come from the three dimensions of collaborative MDSE; we also added a specific sub-question for investigating how collaborative MDSE approaches are integrated with software engineering activities. 


\begin{itemize}

\item \textbf{RQ1.1}: \textit{What are the characteristics of the model management infrastructure of existing collaborative MDSE approaches?}

\item \textbf{RQ1.2}: \textit{What are the characteristics of the collaboration means of existing collaborative MDSE approaches?}

\item \textbf{RQ1.3}: \textit{What are the characteristics of the communication means of existing collaborative MDSE approaches?}


\end{itemize}
  

Objective: to \textit{identify} and \textit{classify} existing collaborative MDSE approaches according to the three dimensions of collaborative MDSE. 

Outcome: a \textit{map} that classifies a set of collaborative MDSE approaches based on different categories (e.g., characteristics of collaborative model editing environments, model versioning mechanisms, model repositories, support for communication and decision making, etc.).

\item \textit{\textbf{RQ2} - What are the challenges of existing collaborative MDSE approaches?}

Objective: to \textit{identify} current limitations and challenges with respect to the state of the art in collaborative MDSE.

Outcome: a \textit{map} that classifies collaborative MDSE approaches with respect to their limitations, faced challenges, and future work.

\item \textbf{RQ3}: \textit{What are the publication trends about collaborative MDSE approaches over time?}
      
%
%

Objective: to \textit{identify} and \textit{classify} the interest of researchers in collaborative MDSE approaches and their various characteristics over time. 


%
%
%
%
%

      
Outcome: a \textit{map} that classifies the collected primary studies according to publication year, venue, applied research strategies, etc. Also, the map will classify collected primary studies according to their focus on the various characteristics of collaborative MDSE approaches over time.

%
%

\end{itemize}

%

The classification resulting from our investigation on RQ1, RQ2, and RQ3 will provide a solid foundation for a thorough identification and comparison of existing and future solutions for supporting collaborative MDSE. This contribution is useful for both \textit{researchers} and \textit{practitioners} willing to further contribute with new collaborative MDSE approaches, or willing to better understand or refine existing ones.

The above listed research questions will drive the whole systematic mapping study methodology, with a special
influence on the primary studies search process, the data extraction process, and the data analysis process.

\section{Search strategy}\label{sec:search}

The success of any systematic study is deeply rooted in the retrieval of a set of primary studies which are relevant and representative enough of the topic being considered \cite{kitchenham2007guidelines}. More specifically, it is fundamental to achieve a good trade-off between the coverage of existing research on the topic considered, and to have a manageable number of studies to be analyzed.
In order to achieve the above mentioned trade-off, as specified in \cite{Babar_identifying}, an optimum search strategy for a SLR should answer the following main questions:

\begin{enumerate}
	\item \textbf{Which} approach to be used in search process?
	\item \textbf{Where} to search, and which part of article should be searched?
	\item \textbf{What} to be searched, and what are queries fed into search engines?
	\item \textbf{When} is the search carried out, and what time span to be searched?
\end{enumerate}

\subsection{Which approaches?}
Our search strategy consists of two main steps: (i) an automatic search on Electronic Data Sources (EDS) and (ii) a snowballing procedure. During the first step, following the guidelines in \cite{kitchenham2007guidelines}, we composed the search string based on identified keywords from research questions and area of study. The search strings are used to retrieve a set of potential primary studies through web search engines provided by digital libraries. The guidelines \cite{kitchenham2007guidelines} recommend to use other complementary searches, necessary to extend the coverage on the topic. For this purpose, we applied a snowballing procedure on the results of the automatic search. Snowballing refers to using the reference list of a paper (backward snowballing) or the citations to the paper (forward snowballing) to identify additional papers \cite{Wohlin14Guidelines}. The \textit{start set} for the snowballing procedure is composed by the selected papers retrieved by automatic search, namely the primary studies selected applying inclusion/exclusion criteria to the automatic search results. In any case, inclusion/exclusion criteria (see Section~\ref{sec:inclusion}) will be applied to each paper and, if the paper can be included, snowballing will be applied iteratively. The procedure ends when no new papers are found.

\subsection{Where to search?}
According to \cite{kitchenham2007guidelines}, it is important to search many different electronic sources, because no single source is able to find all relevant primary studies. Table \ref{tab:engines} shows the electronic databases we will use for our study. These are considered the main sources of literature for potentially relevant studies on software engineering \cite{Babar_EDS}. Also, these EDSs have been selected from the recommendations made by experts in the area of software engineering\footnote{We do not use Google Scholar since it may generate many irrelevant results and have considerable overlap with ACM and IEEE on software engineering literature \cite{Babar_EDS}. However, we will use Google Scholar in the forward snowballing procedure \cite{Wohlin14Guidelines}}.

\begin{center}
	\begin{table}[h]\center
		\begin{tabular}{l | l }
			\hline
			{\bf Library} & {\bf Website} \\ \hline
			ACM Digital Library & \url{http://dl.acm.org} \\
			IEEE Xplore Digital Library & \url{http://ieeexplore.ieee.org} \\
			Web of Science & \url{http://apps.webofknowledge.com} \\
			ScienceDirect &  \url{http://www.sciencedirect.com} \\
			SpringerLink & \url{http://link.springer.com} \\
			Wiley Online Library & \url{http://onlinelibrary.wiley.com/} \\
			\hline
		\end{tabular}
		\caption{Electronic data sources targeted with search strings}
		\label{tab:engines}
	\end{table}
\end{center}

\subsection{What to search?}


A suitable search string will be the input to the electronic data sources identified in the previous section, matching with paper titles, abstracts, and keywords. According to the guidelines provided in \cite{kitchenham2007guidelines}, we will use the following systematic strategy for constructing our search string:
\begin{enumerate}
	\setlength{\itemsep}{1pt}
	\setlength{\parskip}{0pt}
	\setlength{\parsep}{0pt}
	\item derive major aspects relevant to the study, according to the research questions and to a set of known relevant papers \textit{pilots}. Table \ref{tab:pilots} shows the considered pilots. Each aspect is represented by a "cluster" that groups a set of terms; identified clusters are \textit{collaborative} and \textit{MDSE};
	\item add keywords (main terms) to each cluster obtained from known primary studies and research questions;
	\item identify and include in a cluster synonyms and related terms of the main terms;
	\item incorporate alternative spellings and synonyms using Boolean \textit{OR};
	\item link the cluster keywords using Boolean \textit{AND};
\end{enumerate}

\begin{center}
	\begin{table}[h]\center
		\begin{tabular}{ l p{8cm} l } 
			\hline
			{\bf Authors} & {\bf Title} & {\bf Year} \\ \hline
			
			Mar\'oti et al. \cite{maroti2014next} & Next Generation (Meta) Modeling: Web-and Cloud-based Collaborative Tool Infrastructure & 2014 \\
			
			Syriani et al. \cite{syriani2013atompm} & AToMPM: A Web-based Modeling Environment & 2013 \\
			
			Farwick et al. \cite{farwick2010web} & A web-based collaborative metamodeling environment with secure remote model access & 2010 \\ 
			 
			Thum et al. \cite{thum2009slim} & SLIM - A Lightweight Environment for Synchronous Collaborative Modeling & 2009 \\
			
			Cataldo et al. \cite{cataldo2009camel} & CAMEL: a tool for collaborative distributed software design & 2009 \\
			
			Bruegge et al. \cite{bruegge2007unicase} & Unicase- an Ecosystem for Unified Software Engineering Research Tools & 2008 \\
			
			De Lucia et al. \cite{de2007enhancing} & Enhancing collaborative synchronous UML modelling with fine-grained versioning of software artefacts & 2007 \\
			
			Kelly et al. \cite{metaedit+} & Metaedit+: a fully configurable multi-user and multi-tool case and came environment & 1996 \\
			
			\hline
		\end{tabular}
		\caption{Pilots}
		\label{tab:pilots}
	\end{table}
\end{center}

Following this strategy, after a series of test executions and refinements the resulting search string is shown in the listing below.

\lstset{language=Pascal}
\begin{lstlisting}[breaklines=true,backgroundcolor=\color{white},basicstyle=\footnotesize\itshape,captionpos=b,caption=Query string used for automatic studies search]
(collaborat* OR coordinat* OR cooperat* OR concur* OR global)
AND (MDE OR MDD OR MDA OR MDS* OR EMF OR DSL OR DSML OR "model driven" OR "eclipse modeling framework" OR "domain specific language" OR "domain specific modeling language")
\end{lstlisting}

Each electronic data source has a specific syntax for search strings, so we adapted our generic search string to the specific syntax and criteria of each electronic data source.

\subsection{When and what time span to search?}
We will include in our search all the studies coming from the selection step avoiding publication year constraints, so we will not consider publication year as criterion for the search and selection steps.


\section{Selection criteria}\label{sec:inclusion}

As suggested in \cite{kitchenham2007guidelines}, we decided the selection criteria of this study during its protocol definition, so to reduce the likelihood of bias. In the following we provide inclusion and exclusion criteria of our study. In this context, a study will be selected as a primary study if it will satisfy \textit{all} inclusion criteria, and it will be  discarded if it will met \textit{any} exclusion criterion.

\subsection{Inclusion criteria}
\begin{enumerate}[{I}1)]
	\item Studies proposing an MDSE method or technique for supporting the collaborative work of multiple stakeholders on models 
	\item Studies in which models are the primary artifacts within the collaboration process.
	\item Studies providing some kind of validation or evaluation of the proposed method or technique (e.g., via a case study, a survey, experiment, exploitation in industry, formal analysis, example usage).
	\item Studies subject to peer review~\cite{wohlin2012experimentation} (e.g., journal papers, papers published as part of conference proceedings will be considered, whereas white papers will be discarded).
	\item Studies written in English language and available in full-text.
\end{enumerate}

\subsection{Exclusion criteria}
\begin{enumerate}[E1)]
	\item Studies discussing \textit{only} business processes and collaboration practices, without proposing a specific method or technique.
	\item Secondary studies (e.g., systematic literature reviews, surveys, etc.).
	\item Studies in the form of tutorial papers, long abstract papers, poster papers, editorials, because they do not provide enough information.
\end{enumerate}

It is important to note that, even if secondary studies will be excluded (see the E4 exclusion criterion), we will considered them in our study as follows:
\begin{itemize}[-]
	\item for checking the completeness of our set of primary studies (i.e., if any relevant paper will be missing from our study);
	\item for providing a summary of what is already known about collaborative MDSE;
	\item for identifying any important issues to be considered in our study;
	\item for defining what is the contribution of our study to the literature.
\end{itemize}


The definition of the above mentioned criteria has been tested by considering the pilot studies defined in Section~\ref{sec:search}; the criteria have been incrementally refined until they were covering all the pilot studies.

\section{Selection procedure}

As suggested by \cite{wohlin2012experimentation}, two researchers will assess a random sample of the studies, then the inter-researcher agreement will be measured using the Cohen Kappa statistic and reported as a quality assessment of this stage in the final report.
To be successful, the result of the Cohen Kappa statistic must be above or equal to $0.80$, otherwise each disagreement must be discussed and resolved, with the intervention of the team administrator, if necessary.

Moreover, if a primary study is published in more than one paper (for example, if a conference paper is extended to a journal version), only one instance will be counted as a primary study. Mostly, the journal version will be preferred, as it is most complete, but both versions will be used in the data extraction phase~\cite{wohlin2012experimentation}. Moreover, if we will have blocking issues in extracting relevant data, supporting technical reports or communication with authors may also serve as data sources for the extraction~\cite{wohlin2012experimentation}.

\section{Data extraction}\label{sec:extraction}
The goal of this step is to identify and collect from the selected primary studies the appropriate and relevant information to answer our research questions (see section~\ref{sec:rq}). To achieve this goal, we will define a rigorous comparison framework to store the extracted data in a structured manner; the classification framework will be composed of a list of attributes representing the set of data items extracted from each primary study.

The creation of an effective classification framework demands a detailed analysis of the contents of each primary study. In light of this, we will follow a systematic process called \textit{keywording}~\cite{mapping_se} for defining our classification framework.
Basically, keywording aims at reducing the time needed in developing the classification framework and ensuring that  it will take all the primary studies into account~\cite{mapping_se}.

\begin{figure}[h]
	\centering
	\includegraphics[width=.6\columnwidth]{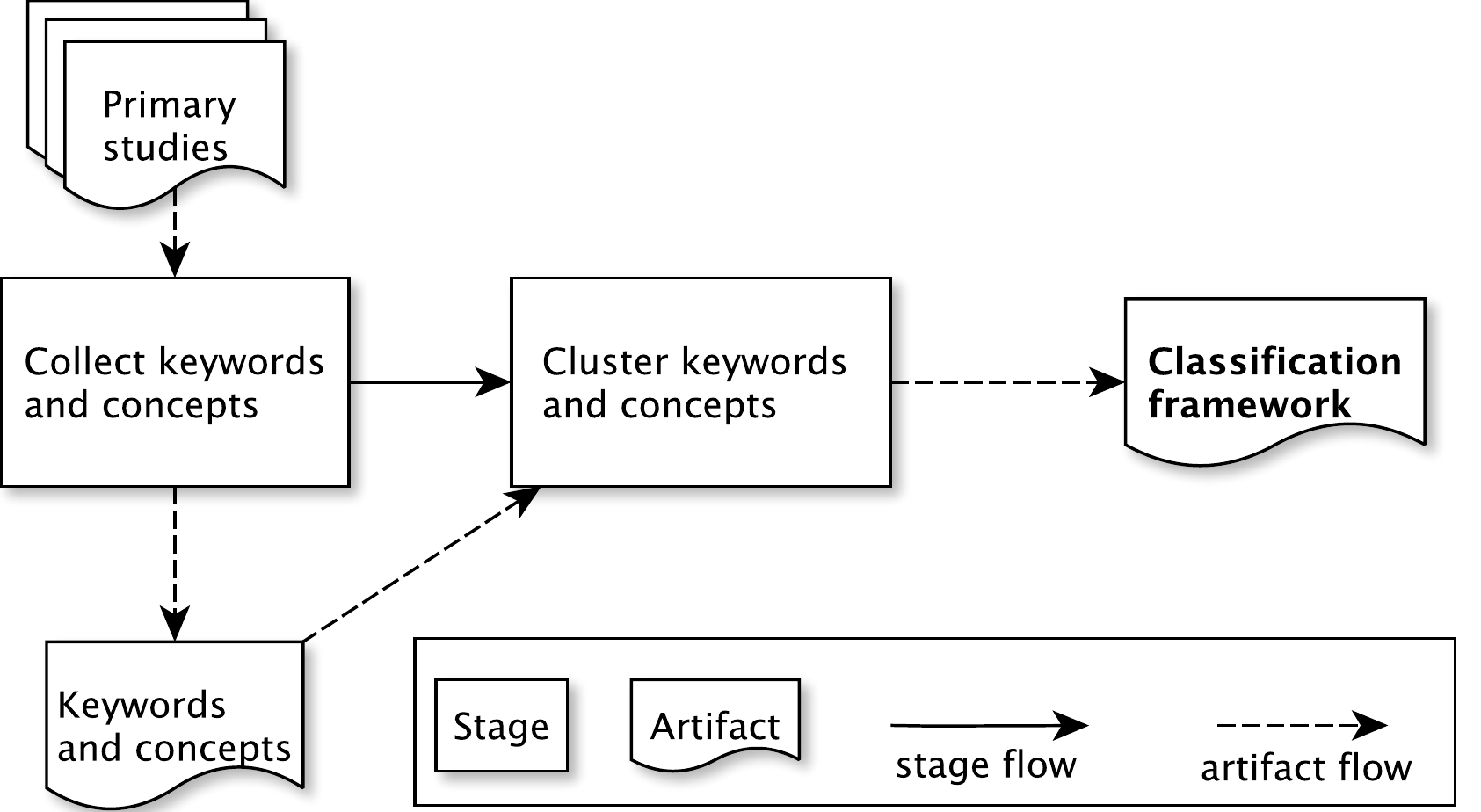}
	\caption{Overview of the keywording process}
	\label{fig:keywording}
\end{figure}

Figure~\ref{fig:keywording} shows the keywording process we will follow.
Keywording is done in two steps:
\begin{enumerate}
	\item \textit{Collect keywords and concepts}: researchers collect keywords and concepts by reading each primary study. When all primary studies have been analyzed, all keywords and concepts are combined together to clearly identify the context, nature, and contribution of the research. The output of this stage is the set of keywords extracted from the primary studies.
	\item \textit{Cluster keywords and concepts}: when keywords and concepts have been finalized, then researchers can perform a clustering operation on them in order to have a set of representative clusters of keywords. The output of this stage is the finalized classification framework containing all the identified attributes, each of them representing a specific aspect regarding collaborative MDSE.
\end{enumerate}

Depending on the resulting classification framework, researchers can decide to extend it with additional attributes that may be of interest in the context of this research. Unless clearly motivated and documented in the final report, attributes cannot be removed from the classification framework.

In order to have a rigorous data extraction process and to ease the management of the extracted data, a structured data\textit{ extraction form} will be designed.
Once the data extraction form will be set up, the principle researcher will consider each primary study and will fill the data extraction form accordingly.
In order to validate our data extraction strategy, we will perform a sensitivity analysis to
analyze whether the results are consistent independently from the researcher performing the analysis~\cite{wohlin2012experimentation}. More specifically, we will get a random sample of 10 primary studies and both the principle and secondary researchers will classify them independently, by filling the data extraction form for each study. Then, the Cohen Kappa statistic will be applied to the obtained results to assess the level of agreement among the researchers. The value of the obtained Cohen Kappa statistics will be documented in the final report, and must be above or equal to $0.80$. If the result of the Cohen Kappa statistic will be below $0.80$, each disagreement must be discussed and resolved, with the intervention of the team advisor, if necessary.

\section{Data synthesis}\label{sec:analysis}

The data synthesis activity involves collating and summarizing the data extracted from the primary studies~\cite[$\S$~6.5]{kitchenham2007guidelines} with the main goal of understanding, analyzing, and classifying
current research on collaborative MDSE. Our data synthesis will be divided into two main phases: vertical analysis, and horizontal analysis.

\noindent \textbf{Vertical analysis}.
We will analyze the extracted data to find trends and collect information about each research question of our study.
Depending on the parameters of the classification framework (see Section~\ref{sec:extraction}), in this research we will apply both quantitative and qualitative synthesis methods, separately.
When considering quantitative data, we will firstly verify if synthesized studies are homogeneously distributed in order to perform meta-analysis. Then, depending on the specific data to be analyzed, we will perform a specific kind of meta-analysis.
When considering qualitative data, we will apply the \textit{line of argument} synthesis~\cite{wohlin2012experimentation}, that is: firstly we will analyze primary studies individually in order to document each of them and tabulate their main features with respect to each specific parameter of the classification framework defined in Section~\ref{sec:extraction}, then we will analyze the set of studies as a whole, in order to reason on potential patterns and trends.
When both quantitative and qualitative analyses will be performed, we will integrate their results in order to explain quantitative results by using qualitative results~\cite[$\S$~6.5]{kitchenham2007guidelines}.

\noindent \textbf{Horizontal analysis}.
In this phase we will analyze the extracted data to explore possible relations across different questions and facets of our research. We will cross-tabulate and group the data, and make comparisons between two or more nominal variables. The main goal of the horizontal analysis is to (i) investigate on the existence of possible interesting relations between data pertaining to different facets of our research. In this context, we will use cross-tabulation as strategy for evaluating the actual existence of those relations.


\section{Dissemination strategy}\label{sec:audience}

We are planning to report our systematic study to different audiences. In the following we list the actions we will undertake in our dissemination strategy:

\begin{enumerate}
	\item we will report our main research-oriented findings and a detailed description of this study into an \textit{academic publication} in a top-level academic journal. Possible targets are: Information and Software Technology journal\footnote{\url{http://www.journals.elsevier.com/information-and-software-technology/}} (IST), Transactions on Software Engineering\footnote{\url{http://www.computer.org/web/tse}} (TSE), or ACM Computing Surveys\footnote{\url{http://csur.acm.org/}} (CSUR);
	\item as suggested in \cite{wohlin2012experimentation}, an accompanying \textit{technical report} will be published on-line; the technical report will present this protocol, the complete lists of included and excluded primary studies, and raw data of this study; the chief aim of the technical report is to make our study replicable by interested researchers;
	\item depending on the nature of the results, we will also target a \textit{practitioners-oriented magazine}, with the goal of (hopefully) impacting and enhancing the current state of the practice of collaborative MDSE.
\end{enumerate}

%

\pagebreak

\bibliographystyle{elsarticle-num}
\bibliography{bibliography}


\end{document}